\documentclass[%
 reprint,
superscriptaddress,
 amsmath,amssymb,
 aps,
floatfix,
]{revtex4-1}

\usepackage{graphicx}
\usepackage{dcolumn}
\usepackage{bm}
\usepackage{amsmath}
\usepackage{xcolor}
\usepackage{comment}
\usepackage[section]{placeins}
\usepackage{float}
\begin{document}

\title{Optical images of massive boson stars with nonlinear electrodynamics}

\author{Xiao-Xiong Zeng}
\affiliation{College of Physics and Electronic Engineering, Chongqing Normal University, Chongqing 401331, China}

\author{Huan Ye}
\affiliation{School of Materials Science and Engineering, Chongqing Jiaotong University, Chongqing 400074, China}

\author{Ke-Jian He}
\affiliation{Department of Mechanics, Chongqing Jiaotong University, Chongqing 400074, China}

\author{Hao Yu}
\email{yuhaocd@cqu.edu.cn, corresponding author}
\affiliation{Physics Department, Chongqing University, Chongqing 401331, China}

\begin{abstract}
This study investigates the optical imaging characteristics of massive boson stars based on a theoretical model with Einstein's nonlinear electrodynamics. Under asymptotically flat boundary conditions, the field equations are solved numerically to obtain the spacetime metric of the massive boson stars. Employing the ray-tracing method, we analyze the optical images of the massive boson stars under two illumination conditions: a celestial light source and a thin accretion disk. The research reveals that the configurations and optical images of the massive boson stars can be tuned via the initial parameter $\phi_0$ and the coupling constant $\Lambda$. The absence of the event horizon in the massive boson stars results in distinct optical image characteristics compared to those of black holes.
\end{abstract}

\maketitle

\section{Introduction}\label{sec:intro}
The study of boson stars initially stemmed entirely from theoretical physicists' curiosity to explore the boundaries between general relativity and fundamental matter theories. A boson star is a hypothetical type of compact astronomical object, unlike a normal star, primarily composed of bosonic particles (like the hypothetical ``axion''~\cite{Kolb:1993zz}) held together by gravity, without relying on fermionic degeneracy pressure~\cite{Kaup:1968zz,Colpi:1986ye,Jetzer:1991jr,Schunck:2003kk,Liebling:2012fv}. In the 1950s, Wheeler first proposed the concept of ``geons'' (something of a ``gravitational atom''), envisioning the existence of quasi-stable structures composed of classical fields of electromagnetism and gravitational field~\cite{Wheeler:1955zz}. However, he found that ``geons'' are difficult to exist stably within the framework of general relativity. It was not until the late 1960s that Kaup introduced a complex scalar field into the theoretical framework, leading to a major breakthrough in this concept~\cite{Kaup:1968zz}. With the complex scalar field, Kaup was able for the first time to theoretically construct stable, star-like solutions, which became the earliest models of boson stars. After decades of development, the study of boson stars has given rise to a variety of theoretical models, exhibiting rich diversity in terms of symmetry, field coupling mechanisms, and dynamical behavior.

With the emergence of ultra-light bosonic dark matter candidates such as axions~\cite{Kolb:1993zz,Eby:2015hsq,Marsh:2015wka}, the foundation was laid for the widespread interests of boson star research. Theoretically, physicists primarily focus on the model construction of boson stars, main characteristics, stability analysis, and so forth. According to the spin of the particles that constitute the boson star, it can be divided into two major categories: scalar boson star~\cite{Kaup:1968zz} and vector boson star (Proca star)~\cite{Brito:2015pxa}. The main characteristics of a boson star include the absence of the event horizon~\cite{Kaup:1968zz,Ruffini:1969qy,Seidel:1990jh,Schunck:2003kk}, no emission of light~\cite{Colpi:1986ye,Sennett:2017etc}, the absence of singularities at the core~\cite{Kaup:1968zz,Ruffini:1969qy}, and other characteristics similar to that of a black hole~\cite{Mielke:2000mh,Guzman:2005bs,Sennett:2017etc}. Its mass can range from tiny to millions of solar masses~\cite{Colpi:1986ye,Raynal:1993qh,Levkov:2016rkk,Dmitriev:2023ipv}. As for stability studies, they serve as the bridge that transforms boson stars from mathematical solutions into physical entities~\cite{Khlopov:1985fch,Gleiser:1988rq,Kusmartsev:1990cr,Liddle:1992fmk,Siemonsen:2020hcg,DiGiovanni:2020frc,Santos:2024vdm}. On the other hand, in terms of observational advancements, the breakthrough in gravitational wave detection and the release of black hole images have ushered in new opportunities for the study of boson stars. As a candidate for dark matter~\cite{Jetzer:1991jr,Lee:1995af,Schunck:1999zu,Schunck:2003kk,Sharma:2008sc,Liebling:2012fv,Marsh:2015wka}, the detection of boson stars has always been a hot topic in cosmology. Since a pair of co-orbiting boson stars is also a potential source of gravitational waves~\cite{Palenzuela:2007dm,Palenzuela:2017kcg,Pacilio:2020jza,Bezares:2022obu}, with the continuous detection of gravitational wave events, some researchers are also analyzing whether certain gravitational wave events, such as GW190521~\cite{CalderonBustillo:2020fyi}, originated from boson stars. Boson stars can serve as the nuclei of galaxies, replacing the supermassive black holes at their centers~\cite{Torres:2000dw,Guzman:2009zz}. Therefore, the existence of boson stars can be determined by detecting the characteristics of massive compact stars at the centers of different galaxies~\cite{Pombo:2023ody,Vincent:2015xta}. Since the Event Horizon Telescope (EHT) Collaboration released images of the compact objects at the center of M87 and Sgr A*~\cite{EventHorizonTelescope:2019dse,EventHorizonTelescope:2022wkp}, there have been new advancements in the detection of boson stars.

This work focuses on the optical properties of massive boson stars within general relativity coupled to nonlinear electrodynamics~\cite{Wang:2023tdz,Yue:2023sep}. The first regular exact black hole solution in general relativity coupled to nonlinear electrodynamics was given by Ayon-Beato and Garciai under the condition of the weak energy~\cite{Ayon-Beato:1998hmi,Ayon-Beato:2000mjt}, which reinterprets the Bardeen model~\cite{J. Bardeen123}
as the gravitational field of a nonlinear magnetic monopole. Nonlinear electrodynamics and the complex scalar field introduce significant modifications to spacetime geometry. We present static, spherically symmetric boson star spacetimes derived from the action, solved numerically under asymptotically flat boundary conditions. We calculate photon geodesics and optical images for two illumination models: a celestial light source and a thin accretion disk. We also investigate how the initial parameter $\phi_0$ and coupling constant $\Lambda$ influence the metric components, photon's effective potential, and optical images of massive boson stars. Recent advances by the EHT in resolving shadow-scale structures of M87* and Sgr A* underscore the urgency of developing discriminative signatures between horizonless objects and black holes~\cite{EventHorizonTelescope:2019dse,EventHorizonTelescope:2022wkp}, which ushers in a new era of research into the images of compact celestial bodies~\cite{Cardoso:2019rvt,Gralla:2019xty,Perlick:2021aok}. Although the related study mainly focuses on the images of black holes~\cite{Vagnozzi:2019apd,Konoplya:2020bxa,Liu:2020ola,He:2022yse,Zeng:2021mok,Pantig:2022gih,Mustafa:2022xod,Guo:2022nto,Chen:2022scf,Gao:2023mjb,Nozari:2023flq,Meng:2024puu,Huang:2024wpj,He:2024amh,Meng:2025ivb,Jafarzade:2025byr}, boson stars has also attracted the attention of many researchers~\cite{Shaymatov:2021nff,Rosa:2023qcv,Zeng:2025xoe,Li:2025awg,Herdeiro:2021lwl,Rosa:2022tfv,Sengo:2024pwk,deSa:2024dhj,He:2025qmq,Zhang:2025xnl}. However, a comprehensive investigation into the optical signatures of boson stars incorporating nonlinear electrodynamics~\cite{Rosa:2022tfv,DeFelice:2024ops,Khoshrangbaf:2025bwg,Rahmatov:2025ari}, especially under realistic accretion scenarios, remains lacking.

The structure of the paper is as follows. Section II develops the theoretical model for constructing a massive boson star in the context of Einstein's nonlinear electrodynamics. In Sections III and IV, we study the optical images of boson stars under celestial light source illumination and thin accretion disk illumination, respectively. Section V presents numerical results and analysis. The last section concludes this work.

\section{Boson Star Solutions}

In this section, we consider a theoretical model for a (massive) boson star combining Einstein's nonlinear electrodynamics with a minimally coupled (complex) scalar field. The action is expressed as~\cite{Ayon-Beato:1998hmi,Ayon-Beato:2000mjt,Wang:2023tdz,Yue:2023sep,J. Bardeen123}
\begin{equation}
	S=\int \sqrt{-g}\, d^4x\left(\frac{R}{4}+\mathcal{L}^{\left(1\right)}+\mathcal{L}^{\left(2\right)}\right )\label{function1},
\end{equation}
where
\begin{equation}
	\mathcal{L}^{\left(1\right)}=-\frac{3}{2s} \frac{\left(2q^2\mathcal{F}\right)^{3/2}}{\left[1+\left(2q^2\mathcal{F}\right)^{3/4}\right]^2} \label{function2},
\end{equation}
\begin{equation}
	\mathcal{L}^{\left(2\right)}=-\nabla_a\Phi^*\nabla^a\Phi-\mu^2\Phi\Phi^*-\Lambda |\Phi|^4 \label{function3}.
\end{equation}
The Lagrangian density $\mathcal{L}^{\left(1\right)}$ is a function of $\mathcal{F}=\frac{1}{4}F_{ab}F^{ab}$ with the electromagnetic field tensor $F_{ab}=\partial_aA_b- \partial_bA_a$. It should be noted that $\mathcal{L}^{\left(1\right)}$ is completely different from the Born-Infeld magnetic field~\cite{Born:1934gh}, and it cannot even reduce to Maxwell electrodynamics. Therefore, the electromagnetic field associated with the Lagrangian density $\mathcal{L}^{\left(1\right)}$ may differ from the one we typically observe on Earth. In this work, we assume that the Lagrangian density of the electromagnetic field (photons) we observe on Earth still takes the standard form of Maxwell electrodynamics. In the Lagrangian density $\mathcal{L}^{\left(2\right)}$, the symbols $\Phi$ and $\Phi^*$ denote the scalar field and its complex conjugate, respectively. The constants $q$ and  $\mu$ represent the magnetic charge and the scalar field mass, respectively. The parameters $s$ and $\Lambda$ are two free  variables.

By varying the action (\ref{function1}) with respect to the metric $g_{ab}$,  the electromagnetic field $A$, and the scalar field $\Phi$, we obtain the following equations of motion:
\begin{equation}
	R_{ab}-\frac{1}{2} g_{ab}R-2\left(T_{ab}^{\left(1\right)}+T_{ab}^{\left(2\right)} \right)=0\label{function4},
\end{equation}
\begin{equation}
	\nabla _a\left(\frac{\partial \mathcal{L}^{\left(1\right)}}{\partial \mathcal{F} }F^{ab} \right)=0\label{function5},
\end{equation}
\begin{equation}
	\Box \Phi -\left(\mu^2+2\Lambda|\Phi|^2\right)\Phi=0\label{function6}.
\end{equation}
Here, $\Box=\nabla^a\nabla_a$ is the d'Alembert operator. $T_{ab}^{(1)}$ and $T_{ab}^{(2)}$ are the energy-momentum tensors of the electromagnetic field and the scalar field, respectively, which are given by
\begin{equation}
	T_{a b}^{\left(1\right)}=-\frac{\partial \mathcal{L}^{\left(1\right)}}{\partial \mathcal{F}} F_{a c} F_{b}{ }^{c}+g_{a b}\, \mathcal{L}^{\left(1\right)},
\end{equation}
\begin{equation}
	\begin{split}
		T_{ab}^{\left(2\right)} &=-g_{ab}\left[\frac{1}{2}g^{ab}(\partial_a\Phi^*\partial_b\Phi+\partial_b\Phi^*\partial_a\Phi)+\mu^2\Phi^*\Phi +\Lambda |\Phi|^4 \right] \\
		&\quad +\partial_a\Phi ^* \partial_b\Phi +\partial_b\Phi ^* \partial_a\Phi.
	\end{split}
\end{equation}

For a general static spherically symmetric solution, the metric can be assumed as
\begin{equation}
	ds^2=-N\left(r\right)\sigma^2\left(r\right)dt^2+\frac{dr^2}{N\left(r\right)}+r^2\left(d\theta^2+sin^2\theta d\varphi^2\right) \label{function9},
\end{equation}
where $N\left(r\right)=1-2m(r)/r$ with  $m(r)$ being a mass-related function, and  $\sigma(r)$ depends entirely on the radial variable $r$. The electromagnetic and scalar fields can  be decomposed as
\begin{equation}
	A=q\, cos\left(\theta\right)d\varphi,~~\Phi=\phi\left(r\right)e^{-i\omega t}\label{function10},
\end{equation}
where $\omega$ is the oscillation frequency of the scalar field, and  $\phi(r)$ is the initial scalar field. Substituting these expressions into the equations of motion yields the following equations for $\phi(r)$, $N(r)$, and $\sigma(r)$:
\begin{equation}
	\phi^{\prime\prime}+\left(\frac{2}{r}+\frac{N^\prime}{N}+\frac{\sigma^\prime}{\sigma}\right)\phi^\prime+\left(\frac{\omega^2}{N\sigma^2}-\mu^2-2\Lambda \phi^2\right)\frac{\phi}{N}=0\label{function11},
\end{equation}
\begin{align}
	N^\prime &+ 2r\left(\mu^2 \phi^2 + \Lambda \phi^4\right) + \frac{2r\omega^2\phi^2}{N\sigma^2} + 2rN\phi^{\prime2} \nonumber+\frac{N}{r} - \frac{1}{r}\\&+ \frac{3q^6r}{s\left(q^3 + r^3\right)^2}  = 0 \label{function12},
\end{align}
\begin{equation}
	\frac{\sigma^\prime}{\sigma}-2r\left(\phi^{\prime2}+\frac{\omega^2\phi^2}{N^2\sigma^2} \right)=0  \label{function13},
\end{equation}
where the notation ``$\prime$'' represents the first derivative with respect to $r$ and ``$\prime\prime$'' denotes the second derivative. These equations constitute a highly nonlinear system. Therefore, when solving these differential equations, it becomes crucial to specify appropriate boundary conditions for each unknown function. In this boson star model, we specify the following asymptotically flat boundary conditions. By requiring the geometry to be asymptotically flat analogous to the Schwarzschild solution, and imposing that the scalar field vanishes as $r\rightarrow\infty$, we have
\begin{equation}
	\left.N\left(\infty\right)=1-\frac{2M}{r}\right|_{r=\infty}=1,~~\sigma\left(\infty\right)=1,~~\phi(\infty)=0,
\end{equation}
{where $ M = m(r \to \infty) $ is the ADM mass.} Moreover, at $ r = 0 $, the solutions of the variables should remain regular. Expanding $ N(r) $, $ \sigma(r) $, and $ \phi(r) $ as power series around $ r = 0 $, one can obtain
\begin{equation}
	N\left(0\right)=1,~~\sigma\left(0\right)=\sigma_0,~~\phi(0)=\phi_0.
\end{equation}
Additionally, for the scalar field \( \Phi \), we have the following specific boundary condition
\begin{equation}
	\left.\frac{d \phi(r)}{d r}\right|_{r=0}=0.
\end{equation}
{In this model, the stellar radius is set to $R=0.98\,M$, as specified in Ref.~\cite{Rosa:2022tfv}.} With the boundary conditions provided above, we can solve the equations of motion for a given $\phi(r)$.

\section{Optical Images Under Celestial Light Source Illumination}
In this section, we discuss the optical image of a boson star illuminated by a celestial light source, approached from the perspective of ray tracing. Since we have assumed that the Lagrangian density of photons we observe takes the standard form of Maxwell electrodynamics, the motion of photons around the boson star is still governed by the Euler-Lagrange equation:
\begin{equation}
	\frac{d }{d\lambda}\left(\frac{\partial\mathcal{L} }{\partial \dot{x}^{\alpha } } \right)=\frac{\partial\mathcal{L} }{\partial x^{\alpha } }.
\end{equation}
Here, $\lambda$ is the affine parameter, $\dot{x}^\alpha$ denotes the photon's four-velocity, and $\mathcal{L}$ is the Lagrangian density of photons. For orbits confined to the equatorial plane $(\theta=\frac{\pi}{2})$, the Lagrangian density $\mathcal{L}$ simplifies to
\begin{equation}
	\begin{split}
		0 = \mathcal{L}
		&= \frac{1}{2} g_{\alpha \beta }\dot{x}^{\alpha}\dot{x}^{\beta} \\
		&= \frac{1}{2}\left[\frac{\dot{r}^{2}}{N(r)} -N(r)\sigma^2(r)\dot{t}^{2}+r^{2}\left(\dot{\theta}^{2}+\sin^{2}\theta\dot{\varphi }^{2}\right)\right],
	\end{split}
	\label{function19}
\end{equation}
where the dot ``$.$'' denotes the first derivative with respect to the affine parameter $\lambda$. Since the metric components exhibit no explicit dependence on time $t$ and azimuthal angle $\varphi$, there are two conserved quantities: the energy $E$ and angular momentum $L$, expressed as
\begin{equation}
	E=-\frac{\partial \mathcal{L} }{\partial \dot{t} } =N(r)\sigma^2(r)\frac{d t}{d \lambda},
	\label{function20}
\end{equation}
\begin{equation}
	L=\frac{\partial \mathcal{L} }{\partial \dot{\varphi } } =r^{2}\frac{d \varphi }{d \lambda }\label{function21}.
\end{equation}
From Eqs.~(\ref{function19}), (\ref{function20}), and (\ref{function21}), we derive the components of the photon's four-velocity along the temporal, azimuthal, and radial directions:
\begin{equation}
	\frac{d t}{d \lambda} =\frac{E}{ N(r)\sigma^2(r)}
	\label{function22},
\end{equation}
\begin{equation}
	\frac{d \varphi }{d  \lambda } =\pm \frac{L}{r^{2}}
	\label{function23},
\end{equation}
\begin{equation}
	\frac{d r}{d  \lambda } =L\sqrt{\frac{1}{h^{2}}\frac{1}{\sigma^2(r)}-\frac{1}{r^{2}} N(r)}
	\label{function24}.
\end{equation}
The sign ``$+$'' denotes clockwise motion and ``$-$'' denotes counterclockwise motion. The parameter $ h $ is defined as
\begin{equation}
	h=\frac{ L }{E},
\end{equation}
representing the ratio of the angular momentum to the energy, also known as the impact parameter. Consequently, the photon's equation of motion can be expressed as
\begin{equation}
	\frac{1}{N(r)}\dot r^2+V_{eff}(r)=0,
\end{equation}
where the (photon's) effective potential \( V_{eff}(r) \) governing the geodesics is given by
\begin{equation}
	V_{eff}(r)=-\frac{E^2}{N\,\sigma^2}+\frac{L^2}{r^2}
	\label{function27}.
\end{equation}
The effective potential is directly related to the existence of photon rings. Specifically, if the effective potential has extrema, then the boson star possesses photon rings, and the locations of these rings are determined by the first derivative of the effective potential, i.e., $V^{\prime}_{eff}(r) = 0$. Furthermore, the stability of the photon orbits around a boson star can be determined by the second derivative of the effective potential, i.e.,
\begin{equation}
	V_{eff}''(r) \left\{
	\begin{array}{ll}
		>0, &~~ \text{Stability} \\
		=0, & ~~\text{Marginal stability} \\
		<0, & ~~\text{Unstability}
	\end{array}
	\right. \label{function28}.
\end{equation}

To determine the motion of photons, the geodesic equation alone is insufficient. We need to specify the initial conditions for the evolution. For this purpose, we select a specific observer in the boson star spacetime.{ We assume that a zero-angular-momentum observer (ZAMO) is located at the point $O(t_o,~r_o,~\theta_o,~\varphi_o)$.} In the vicinity of the ZAMO, there exists a locally orthonormal tetrad, which is given by
\begin{equation}
	\begin{split}
		\hat{e}_{0} &= \frac{1}{\sqrt{-g_{tt}}} \partial_t, \quad \hat{e}_{1} = -\frac{1}{\sqrt{g_{rr}}} \partial_r, \\
		\hat{e}_{2} &= \frac{1}{\sqrt{g_{\theta\theta}}} \partial_\theta, \quad \hat{e}_{3} = -\frac{1}{\sqrt{g_{\varphi \varphi}}} \partial_\varphi.
	\end{split}
\end{equation}
In the ZAMO's coordinate system, the tangent vector of a null geodesic $v_o$ can be expressed as
\begin{equation}
	\dot{v}_o=\mid \overrightarrow{OF}\mid  (-\hat{e}_{0}+\cos \Omega \,\hat{e}_{1}+\sin \Omega \cos \zeta\, \hat{e}_{2}++\sin \Omega \sin \zeta \,\hat{e}_{2}),\label{function30}
\end{equation}
where $\overrightarrow{OF}$ is the tangent vector of the null geodesic at the point $O$ in the three-dimensional subspace, $\Omega$ denotes the angle between $\overrightarrow{OF} $ and $\hat{e}_{1}$, and  $\zeta$ denotes the angle between $\overrightarrow{OF} $ and $\hat{e}_{2}$. The so called celestial coordinate system is described by $(\Omega,\zeta)$~\cite{Li:2024ctu}. On the other hand, for a null geodesic $v(\alpha)$ expressed as the coordinates $t(\alpha)$, $r(\alpha)$, $\theta (\alpha)$, and $\varphi(\alpha)$, the general form of the tangent vector is given by
\begin{equation}
	\dot{v}(\alpha)=\dot{t}\partial_t+\dot{r}\partial_r+\dot{\theta}\partial_{\theta}+\dot{\varphi}\partial_{\varphi}.\label{function31}
\end{equation}
Comparing Eqs. (\ref{function30}) and (\ref{function31}), we conclude that once the photon's four-momentum is fixed, the celestial coordinates are determined. Conversely, if the celestial coordinates are known, the photon's four-momentum can be obtained through a coordinate transformation. Therefore, combined with the ZAMO's position, the initial conditions for the photon's equation of motion can be directly derived. To generate the optical image of the boson star, the celestial coordinates $(\Omega,\zeta)$ must be mapped one-to-one to the points $(x,y)$ on the image plane. If the image plane is divided into $\mathfrak n$ equal-sized squares, then for any point with pixel coordinates $(i,j)$, the corresponding celestial coordinates can be expressed as
\begin{equation}
	\tan \zeta =\frac{2j -(\mathfrak n+1)}{2i -(\mathfrak n+1)},
\end{equation}
\begin{equation}
	\tan \Omega =\frac{1}{\mathfrak n} \tan\frac{1}{2} \gamma_{_{fov}}\sqrt{\left(i-\frac{\mathfrak n+1}{2}\right)^{2}+\left(j-\frac{\mathfrak n+1}{2}\right)^2},
\end{equation}
{where $\gamma_{_{fov}}$ represents the angle of field of view.} Once the pixel resolution and the field angle $\gamma_{_{fov}}$ are determined, along with the celestial coordinates and initial conditions, the photon's trajectory in the boson star spacetime can be computed by solving the photon's equation of motion.

\section{Optical Image Under Thin Accretion Disk Illumination}
In this section, we will investigate the optical image of a boson star illuminated by a thin accretion disk. Here, the accretion disk is geometrically and optically thin, and is fixed on the equatorial plane of the boson star. When light interacts with the thin accretion disk, its intensity is altered by both emission and absorption processes. Assuming negligible refraction effects, the variation in light intensity follows the form
\begin{equation}
	\frac{d}{d \lambda}\left(\frac{Q_{\nu}}{\nu^3}\right)=\frac{\mathcal J_{\nu}-k_{\nu} Q_{\nu}}{\nu^2}\label{function34},
\end{equation}
where $\lambda$ is still the affine parameter of the null geodesic. The variables $Q_{\nu}$, $\mathcal J_{\nu}$, and $k_{\nu}$ represent the specific intensity, emission coefficient, and absorption coefficient at frequency $\nu$, respectively. When light propagates in a vacuum, $\mathcal J_{\nu}$ and $k_{\nu}$ vanish, and thus ${Q_{\nu}}/{\nu^3}$ remains conserved along the geodesic.

Under the thin disk approximation, we only consider instantaneous emission and absorption within the equatorial plane, meaning that $\mathcal J_{\nu}$ and $k_{\nu}$ exist solely on the equatorial plane and vanish elsewhere. {In this case, the total specific intensity observed by the ZAMO} can be written as
\begin{equation}
	Q_{0}=\sum\limits_{n}^{n_{max}} f_n\left(\frac{\nu_0}{\nu_n}\right)^3J_n \label{function35},
\end{equation}
where $n=1...n_{max}$ is the number of times the light passes through the equatorial plane. The undetermined variables $f_n$ and $J_n$ are the ``fudge factor'' and a variable related to the emission of the accretion disk, respectively. Here, {$\nu_0$ is the observed frequency on the screen, $\nu_n$ is the frequency observed by the local rest frame comoving with the accretion disk}. We define $g_n \equiv \frac{\nu_0}{\nu_n} $ as the redshift factor.

To determine the total specific intensity $Q_{0}$, we must first specify the exact forms of $f_n$, $J_n$, and $g_n$. Since $f_n$ primarily modifies the intensity of photon rings (with limited impact on the overall image), we can set it to 1 for simplicity~\cite{Hou:2022eev,Yang:2024nin}. For the emission variable $J_n$, multiple functional forms are possible. To align with astronomical observations, such as those of M87* and Sgr A*, it is commonly modeled as a second-order polynomial in logarithmic space. In this work, we adopt the Gralla-Lupsasca-Marrone model~\cite{Gralla:2020srx}, where the accretion disk emission profile is given by
\begin{equation}
	J=\frac{\exp \left[-\frac{1}{2} \left(\epsilon +\sinh ^{-1}\left(\frac{r-\xi }{\eta }\right)\right)^2\right]}{\sqrt{\left(r-\xi \right)^2+\eta ^2}}.\label{function36}
\end{equation}
This model's predictions align with magnetohydrodynamic simulations of accretion disks in general relativistic astrophysics~\cite{Vincent:2022fwj}. The parameters $\epsilon$, $\xi$, and $\eta$ control the shape of the emission profile and are termed the growth rate, radial shift, and profile dilation, respectively. In principle, these parameters can be adjusted to select an appropriate intensity distribution range for the studied model. For the sake of computational simplicity, we adopt the parameter values  $\epsilon=0$, $\xi=6M$, and $\eta=M$. In this model, the accretion flow consists of electrically neutral plasma moving along time-like geodesics with conserved quantities $E$ (energy), $L$ (angular momentum), and $\Omega_{n}$ (angular velocity). The specific expression for the angular velocity $\Omega_{n}$ is given by
\begin{equation}
	\Omega_{n}=\left.\frac{u^{\varphi}}{u^t } \right|_{r=r_n},
\end{equation}
where $n = 1...n_{max}$ still denotes the number of times the light passes through the equatorial plane. Under these conditions,  the redshift factor is given by~\cite{Hou:2022eev}
\begin{equation}
	g_{n}=\frac{1}{\Sigma \left (1-h \,\Omega_{n}\right)} \label{function38},
\end{equation}
where $\Sigma$ is defined as
\begin{equation}
	\Sigma =\left.\sqrt{\frac{-1}{g_{tt}+g_{\varphi\varphi}\Omega_{n}^2}} \right|_{r=r_n} \label{function39}.
\end{equation}
Based on the above equations, one can find that in order to obtain the optical image of a boson star illuminated by a thin accretion disk, it is necessary to take into account both the light intensity and the redshift effect.

\section{Numerical Results}
With the theoretical framework established, we now proceed to obtain boson star solutions using the numerical method and further investigate the optical images of boson stars under different background light sources. In this model, the boson star solutions are characterized by several parameters: the initial scalar field $\phi$, magnetic charge $q$, free parameter $s$, and coupling constant $\Lambda$. Our study primarily focuses on how variations in the initial scalar field and coupling constant influence the optical images of boson stars. To begin, we fix the values of the magnetic charge $q$, free parameter $s$, and coupling constant $\Lambda$, and examine how changes in the initial scalar field $\phi$ affect both the configurations of boson stars and the corresponding optical images. We then fix the magnetic charge $q$, free parameter $s$, and initial scalar field $\phi$ to explore  the influence of the coupling constant $\Lambda$ on these two aspects.

\begin{figure}[H]
	\centering
	\includegraphics[scale=0.7]{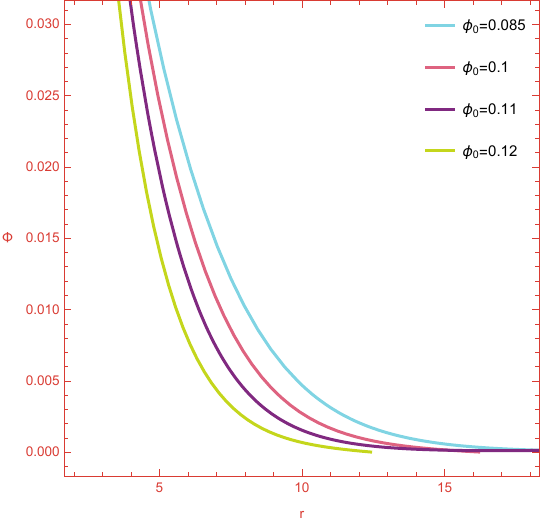}
	\caption{Variation of the scalar field $\Phi(r)$ with respect to the radial coordinate $r$ for different values of $\phi_0$. The values of $\phi_0$ considered are 0.085, 0.1, 0.11, and 0.12.}\label{fig1}
\end{figure}

\subsection{Influence of the Initial Parameter $\phi_0$ on Optical Images}
In this section, the values of the magnetic charge $q$, free parameter $s$, and coupling constant $\Lambda$ are fixed at $q = 1$, $s = 1$, and $\Lambda = 100$, respectively. For other values of these parameters, the results are similar. We first numerically solve the scalar field and metric components, and then obtain the fitting functions for the metric components through curve fitting. Based on the fitted metric, we can further investigate the optical images of boson stars surrounded by a celestial light sources and a thin accretion disk.

\begin{figure}[H]
	\centering
	\includegraphics[scale=0.40]{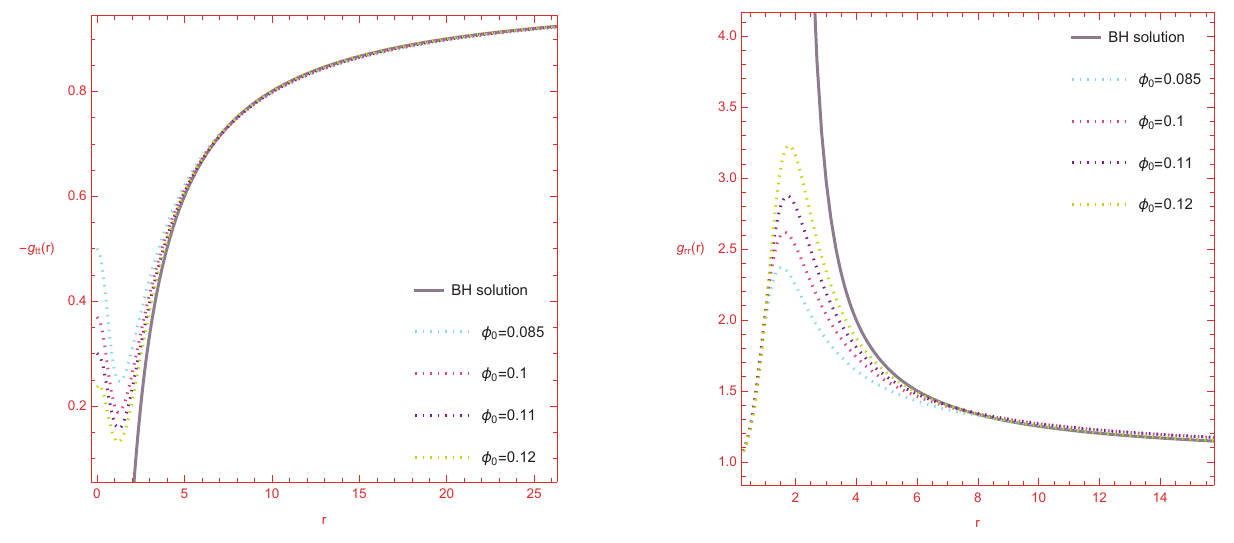}
	\caption{Comparison of the metric components $-g_{tt}$ (left panel) and $g_{rr}$ (right panel) between boson stars and the Schwarzschild black hole for different values of $\phi_0$. The dark solid lines indicate the Schwarzschild black hole, while the dashed lines represent boson stars. The values of $\phi_0$ considered are 0.085, 0.1, 0.11, and 0.12.}\label{fig2}
\end{figure}

The numerical results of the scalar field $\Phi(r)$ for different values of the initial scalar field $\phi(r)$ at $r=0$ (denoted as the initial parameter $\phi_0$) are shown in Fig.~\ref{fig1}. Given the direct correspondence between $\phi$ and $\phi_0$, for convenience, we shall henceforth use $\phi_0$ in place of $\phi$ to study the influence of the initial scalar field. It is observed that for all values of $\phi_0$, the scalar field remains confined to a narrow region near $r=0$, consistent with all boson star models. Specifically, the value of the scalar field decreases rapidly with increasing radial coordinate and approaches zero as $r\rightarrow\infty$. The metric components $-g_{tt}$ and $g_{rr}$, for different values of $\phi_0$, are illustrated in Fig.~\ref{fig2}. To compare with the simplest spherically symmetric black hole, the metric components of the Schwarzschild black hole with the same mass parameter are also plotted in Fig.~\ref{fig2}, depicted as dark solid lines. It can be found that the Schwarzschild black hole diverges at the event horizon. Unlike black holes, boson stars lack the event horizon, and consequently, their metric components do not exhibit divergence, as evidenced by Fig.~\ref{fig2}. The asymptotic behavior of boson stars is identical to that of the Schwarzschild black hole, both displaying asymptotic flatness, that is, their metric components approach 1 as $r$ increases. 

After obtaining the boson star solutions, we employ numerical simulations to study their optical images. However, due to the presence of numerical infinity, we can assume specific functions to fit the metric components, thereby facilitating subsequent calculations. Drawing on Ref.~\cite{Rosa:2022tfv}, we adopt  the following functions to fit the metric components:
\begin{equation}
	g_{tt}=-\exp\left[\alpha_7 \left(\exp \left(-\frac {1+\alpha_1 r+\alpha_2 r^2}{\alpha_3+\alpha_4 r+\alpha_5r^2+\alpha_6r^3}\right)-1\right)\right],
	\label{function40}
\end{equation}
\begin{equation}
	g_{rr}=\exp\left[\beta_7 \left(\exp \left(-\frac {1+\beta_1 r+\beta_2 r^2}{\beta_3+\beta_4 r+\beta_5r^2+\beta_6r^3}\right)-1\right)\right].
	\label{function41}
\end{equation}

\begin{figure}[H]
	\centering
	\includegraphics[scale=0.40]{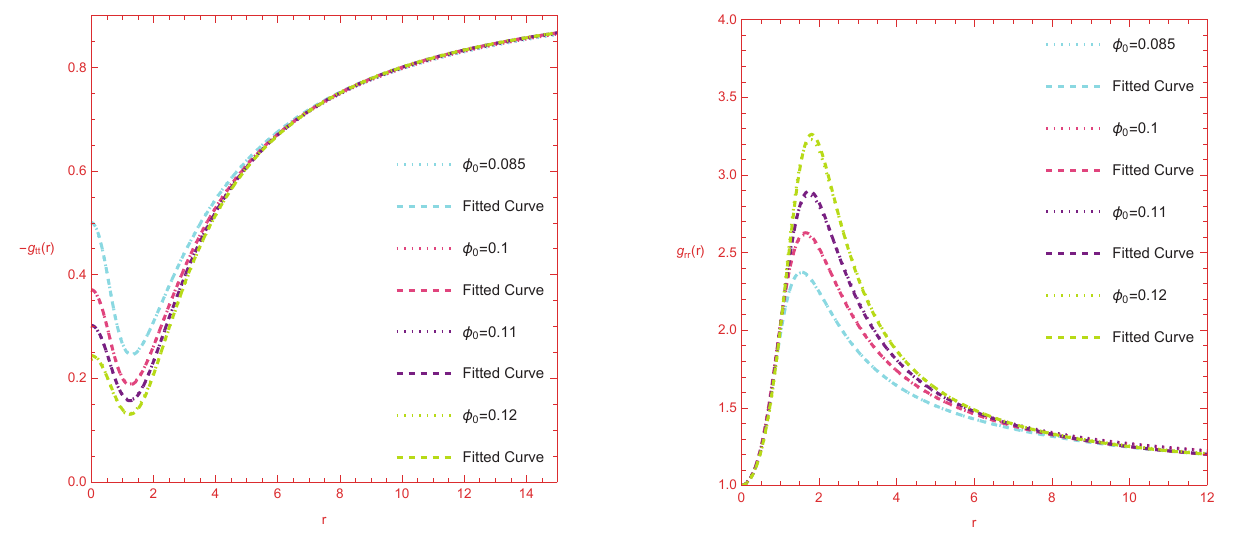}
	\caption{Comparison of the metric components $-g_{tt}$ (left panel) and $g_{rr}$ (right panel) between the numerical and fitted results for different values of $\phi_0$. The numerical results are depicted with the dotted lines, while the fitted results are depicted with the dashed lines. The values of $\phi_0$ considered are 0.085, 0.1, 0.11, and 0.12.}\label{fig3}
\end{figure}

We present the numerical and fitted results for the metric components $-g_{tt}$ and $g_{rr}$ in Fig.~\ref{fig3}. It is evident that, for all values of $\phi_0$ considered, the numerical results and the fitted curves exhibit excellent agreement. The parameters of the fitting functions, along with the boson star's ADM mass, are provided in Tables~\ref{tab1} (for $-g_{tt}$) and~\ref{tab2} (for $g_{rr}$).

\begin{table}[H]
	\centering
	\caption{Values of $\alpha_i$ in the fitted metric component $-g_{tt}$ [see Eq.~(\ref{function40})] for different values of $\phi_0$ and $M$.}\label{tab1}
	\vspace{0.5mm}
	\begin{tabular}{ccccc}
		\hline
		Type&$M_{\phi_0}BS1$& $M_{\phi_0}BS2$ &$M_{\phi_0}BS3$  & $M_{\phi_0}BS4$\\
		\hline
		$\phi_0$ &0.085  &0.1  &0.11 &0.12   \\
		$M$ & 0.908097 &0.90825 &0.879495  &0.847688 \\
		$\alpha_1$&-0.707 &-0.615 &-0.593  &-0.118   \\
		$\alpha_2$& 1.273 &1.008  &0.838   &0.945\\
		$\alpha_3$&-0.553 &-0.541 &-0.525  &-91372.7 \\
		$\alpha_4$&0.4    & 0.325 &0.301   &-18455 \\
		$\alpha_5$&-0.402 &-0.289 & -0.222 &64945.9 \\
		$\alpha_6$&-0.095 &-0.103 &-0.1    &-73573.4\\
		$\alpha_7$&-0.136 &-0.186 &-0.21   &-131973 \\
		\hline
	\end{tabular}
\end{table}

\begin{table}[H]
	\centering
	\caption{Values of $\beta_i$ in the fitted metric component $g_{rr}$ [see Eq.~(\ref{function41})] for different values of $\phi_0$ and $M$.}\label{tab2}
	\vspace{0.5mm}
	\begin{tabular}{ccccc}
		\hline
		Type&$M_{\phi_0}BS1$& $M_{\phi_0}BS2$ &$M_{\phi_0}BS3$  & $M_{\phi_0}BS4$\\
		\hline
		$\phi_0$ &0.085  &0.1  &0.11 &0.12   \\
		$M$ & 0.908097 &0.90825 &0.879495 &0.847688 \\
		$\beta_1$&-17.408 &-20.092 & -18.104 &-13.239   \\
		$\beta_2$&-30.123 &-11.318 &-6.495   &-44.908\\
		$\beta_3$&18.259  &5.37    &5.158    &144.984 \\
		$\beta_4$&-12.799 &-0.308  &-0.536   &-61.967 \\
		$\beta_5$&20.849  &5.625   &4.17     &15.69 \\
		$\beta_6$&2.104   &0.245   &0.151    &45.815\\
		$\beta_7$&-0.18   &-0.052  &-0.064   &-1.827 \\\hline
	\end{tabular}
\end{table}

\begin{figure}[H]
	\centering
	\includegraphics[scale=0.55]{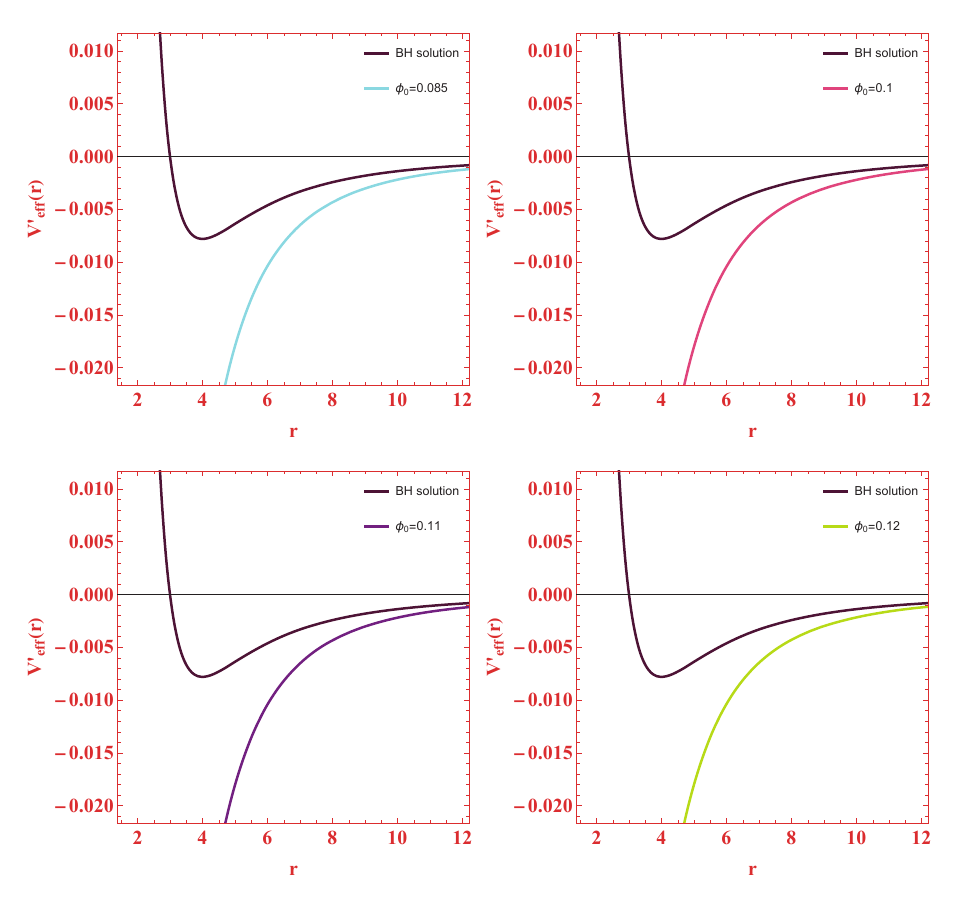}
	\caption{The first derivative of the effective potential for the boson stars ($M_{\phi_0}BS1-M_{\phi_0}BS4$) and the Schwarzschild black holes with the same mass parameters for different values of $\phi_0$. From left to right, in the first row $\phi_0=0.085$ and 0.1, and in the second row $\phi_0=0.11$ and 0.12.}\label{fig4}
\end{figure}

To analyze the photon rings of boson stars, according to Eq.~(\ref{function27}), we plot the first derivative $V^{\prime}_{eff}(r)$ of the effective potential for the boson stars ($M_{\phi_0}BS1-M_{\phi_0}BS4$) in Fig.~\ref{fig4}. For comparison with black holes, each figure includes a dark solid line representing the first derivative of the effective potential for the Schwarzschild black hole with the same mass parameter. The results demonstrate that $V^{\prime}_{eff}(r)$ for the boson stars increases monotonically with the radial coordinate $r$. As $r\rightarrow \infty$, $V^{\prime}_{eff}(r)$ approaches zero but does not intersect the horizontal axis. This implies that the equation $V^{\prime}_{eff}(r) = 0$ possesses no real solutions, consequently confirming the absence of photon rings in the boson star models ($M_{\phi_0}BS1-M_{\phi_0}BS4$). In contrast to the boson stars, the first derivative of the effective potential for the Schwarzschild black holes vanishes at $r = 3M$, corresponding to the locations of the photon rings of the Schwarzschild black holes. Consequently, the optical signatures of the boson stars differ distinctly from those of the Schwarzschild black holes due to the absence of photon rings.

\begin{figure}[H]
	\centering
	\includegraphics[scale=0.55]{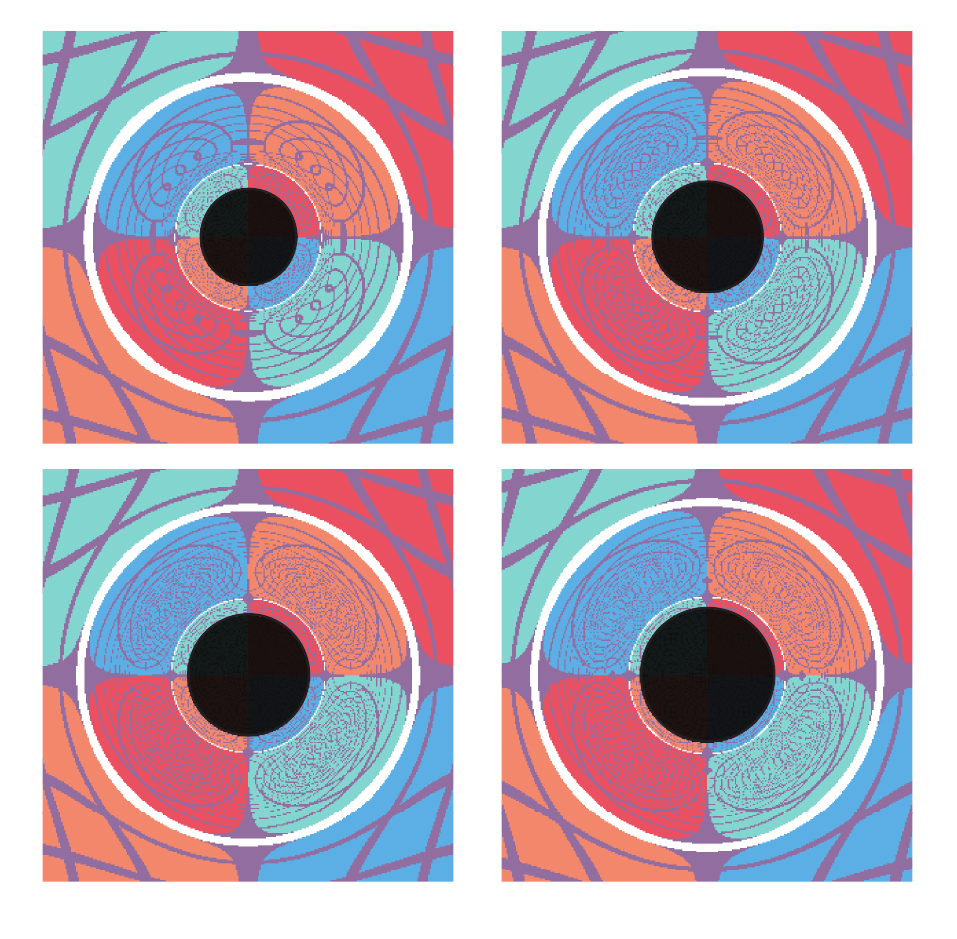}
	\caption{Optical images of boson stars and their Einstein rings observed at an observation inclination angle of $\theta=75^\circ$ with a field angle of $\gamma_{_{fov}}=18^\circ$ under celestial light source illumination. From left to right, in the first row $\phi_0=0.085$ and 0.1, and in the second row $\phi_0=0.11$ and 0.12. The magnetic charge $q$, free parameter $s$, and coupling constant $\Lambda$ are fixed at $q=1$, $s=1$, and $\Lambda=100$, respectively.}\label{fig5}
\end{figure}

In Fig.~\ref{fig5}, we plot the optical images of the boson stars  ($M_{\phi_0}BS1-M_{\phi_0}BS4$) and their Einstein rings under celestial light source illumination. The observation inclination angle (the angle between the observer's line of sight and the boson star's central axis~\cite{Lee:2022rtg,He:2024qka}) is set to $\theta=75^\circ$, and the field angle (the angle between a point in space and the projection screen~\cite{Guo:2024mij}) is set to $\gamma_{_{fov}}=18^\circ$. In each panel, the celestial sphere is divided into four distinct quadrants, with each sector delineated by a distinct color~\cite{Huang:2024gtu}. Both longitude and latitude grid lines are represented by brown lines, spaced at intervals of $3.75^\circ$. The observer is positioned at the intersection of the four quadrants. On the opposite side of the celestial sphere relative to the observer, a white reference light source is placed to study the strong gravitational lensing effects and the formation of Einstein rings. The central black translucent region outlines the shape of the boson star, while the outer white circle marks the Einstein ring. These images reveal several intriguing phenomena. Unlike black holes, due to the absence of the event horizon, light entering the boson star is not entirely absorbed, and so some photons escape from the interior of the boson star. This explains the presence of colored features within the central region of the boson star in Fig.~\ref{fig5}. To delineate the boson star's size and shape, the central region is rendered in light black, with parameters defined by the ADM mass and stellar radius. On the other hand, similar to black holes, boson stars bend light via a gravitational lens, producing Einstein rings. Since these boson stars are non-rotating, their shape remains spherical, and the central light-black region appears as a perfect circle. These pictures also reveal how changes in the initial parameter $\phi_0$ affect the optical images of the boson stars. Results show that increasing $\phi_0$ enlarges visibly both the boson star's radius and the Einstein ring's radius. This occurs because a stronger initial scalar field intensifies the gravitational field, amplifying light deflection and thus expanding the Einstein ring.

\begin{figure}[H]
	\centering
	\includegraphics[scale=0.4]{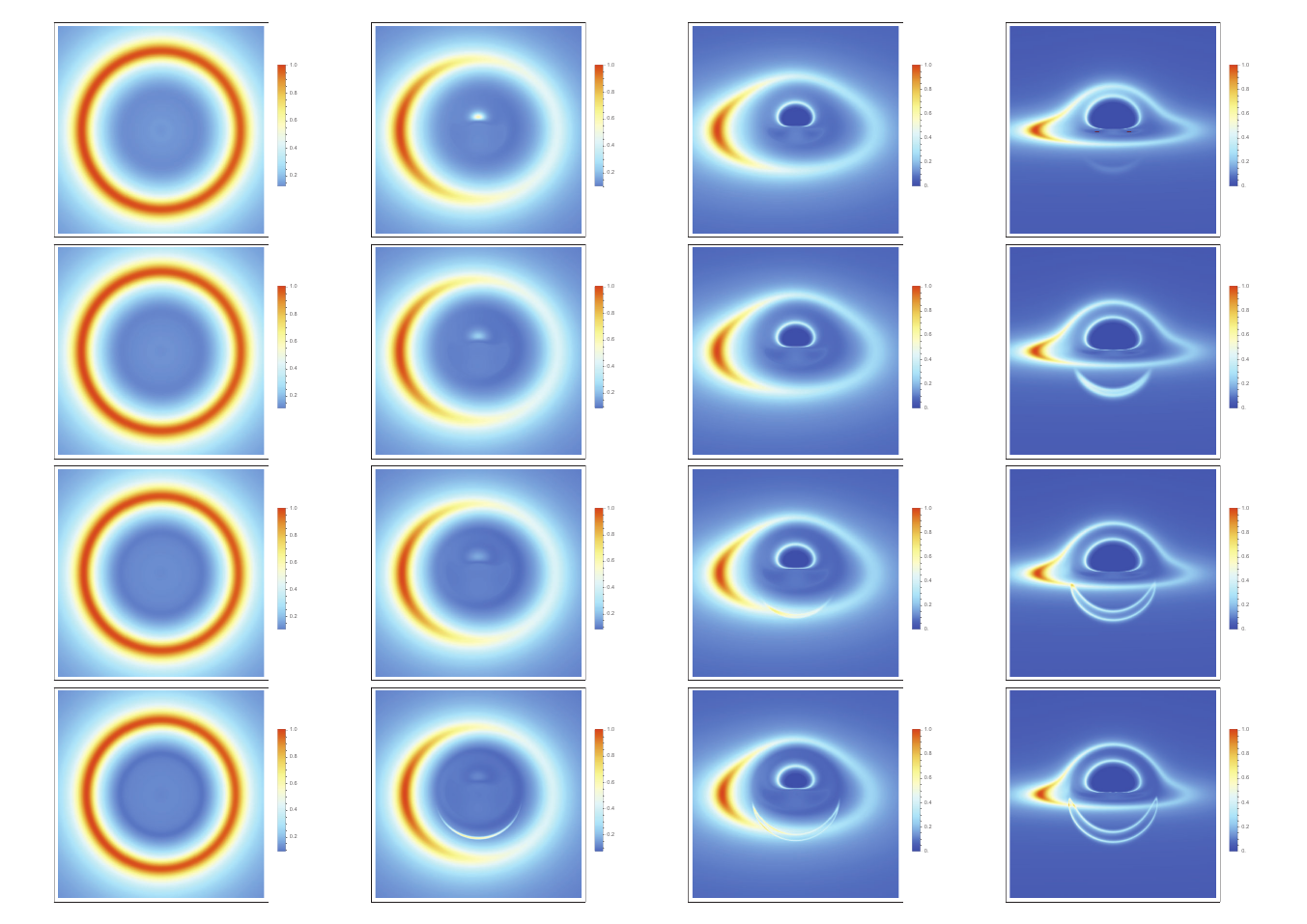}
	\caption{Optical images of boson stars displayed with a field angle $\gamma_{_{fov}}=5^\circ$ under thin accretion disk illumination. The vertical arrangement, from top to bottom, corresponds to $M_{\phi_0}BS1$ ($\phi_0=0.085$), $M_{\phi_0}BS2$ ($\phi_0=0.1$), $M_{\phi_0}BS3$ ($\phi_0=0.11$), and $M_{\phi_0}BS4$ ($\phi_0=0.12$), respectively. Horizontally, the values of the observation inclination angle $\theta$, from left to right, are $1^\circ$, $25^\circ$, $60^\circ$, and $80^\circ$, respectively. The magnetic charge $q$, free parameter $s$, and coupling constant $\Lambda$ are fixed at $q=1$, $s=1$, and $\Lambda=100$, respectively.}\label{fig6}
\end{figure}

When considering the background light source as a thin accretion disk model, the images of the boson stars ($M_{\phi_0}BS1-M_{\phi_0}BS4$) are shown in Fig.~\ref{fig6}. All images display a bright ``ring-like'' structure, with a darker region inside the bright ring. The contour of this dark region represents the optical observational shape of the boson star. For small observation inclination angles (for example, the first column), only the direct image of the accretion disk appears, where $n=1$ (indicating that photons cross the equatorial plane once), and the bright ring appears as an axisymmetric circular ring. As $\phi_0$ increases (from top to bottom), the size of the bright ring gradually decreases. As $\theta$ increases (from left to right), the bright ring evolves from a ``disk-like'' shape to a ``hat-like'' shape, and a distinct lensed image appears, where $n=2$ (indicating that photons cross the equatorial plane twice). The size of the lensed image also increases with $\theta$. In Fig.~\ref{fig4}, the first derivative of the effective potential for the boson stars ($M_{\phi_0}BS1-M_{\phi_0}BS4$) does not intersect the horizontal axis, meaning that none of these boson stars possess photon rings. This result is fully consistent with Fig.~\ref{fig6}, which similarly shows no evidence of photon rings.

From Fig.~\ref{fig6}, it can also be seen that a smaller bright ring structure appears inside the central dark region. As $\theta$ increases, this bright ring gradually enlarges, and the distribution of maximum light intensity (the illuminated part in the image) gradually shifts toward the left side, while the light intensity on the right side shows a diminishing trend. This arises from the Doppler redshift effect, which becomes more pronounced on the right side as $\theta$ increases.

\begin{figure}[H]
	\centering
	\includegraphics[scale=0.4]{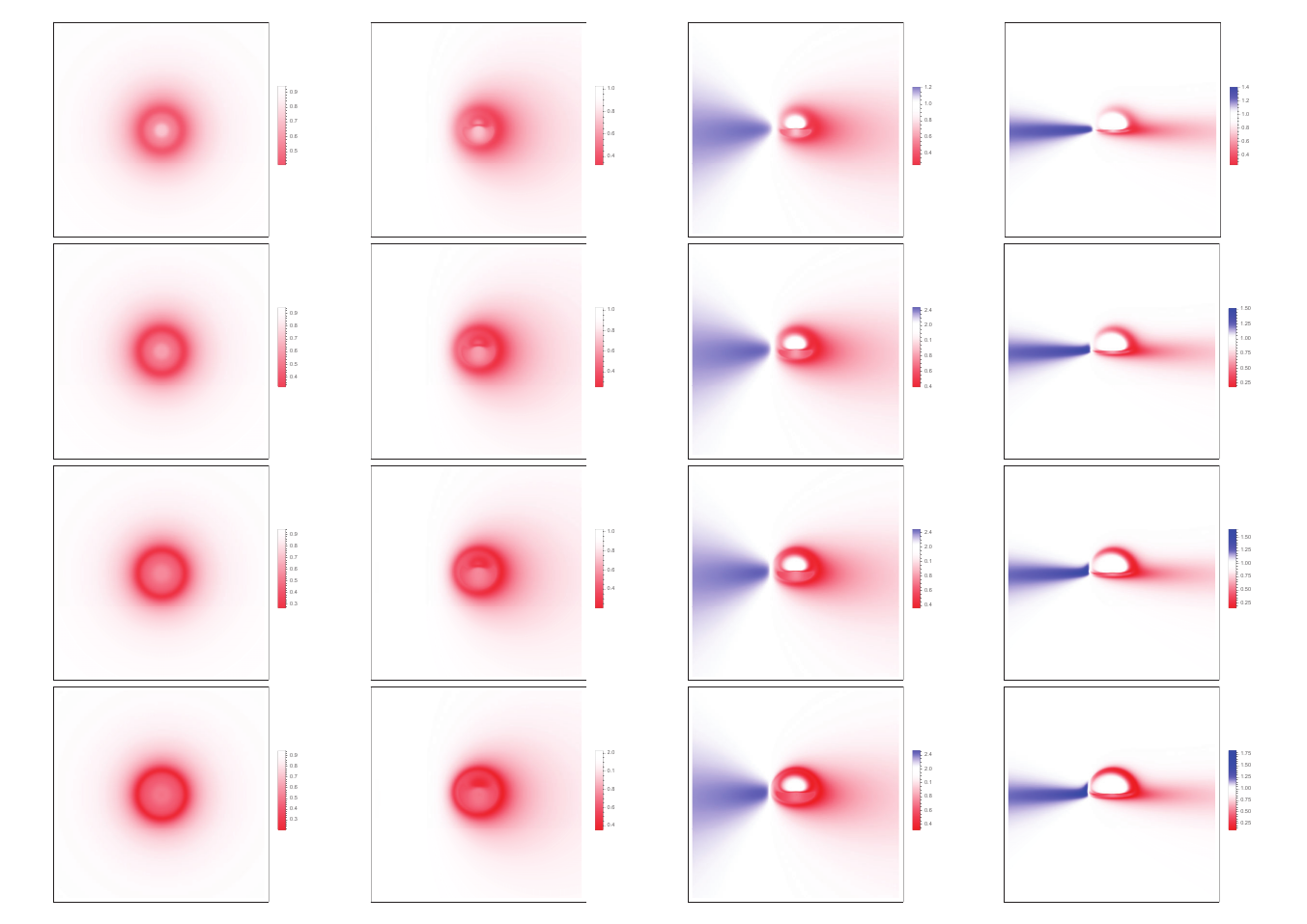}
	\caption{Images of the redshift and blueshift effects of a thin accretion disk with a field angle $\gamma_{_{fov}}= 10^\circ$. The redshift factor distribution depicted in red and the blueshift factor distribution depicted in blue. The vertical arrangement, from top to bottom, corresponds to $M_{\phi_0}BS1$ ($\phi_0=0.085$), $M_{\phi_0}BS2$ ($\phi_0=0.1$), $M_{\phi_0}BS3$ ($\phi_0=0.11$), and $M_{\phi_0}BS4$ ($\phi_0=0.12$), respectively. Horizontally, the values of the observation inclination angle $\theta$, from left to right, are $1^\circ$, $25^\circ$, $60^\circ$, and $80^\circ$, respectively. The magnetic charge $q$, free parameter $s$, and coupling constant $\Lambda$ are fixed at $q=1$, $s=1$, and $\Lambda=100$, respectively.}\label{fig7}
\end{figure}

The images illustrating the redshift and blueshift effects of a thin accretion disk are shown in Fig.~\ref{fig7}. According to Eq.~(\ref{function35}), when $\theta=1^\circ$, the redshift manifests in the direct image, and due to the small observation inclination angle, the primary observed effect is the gravitational redshift. Moreover, as $\theta$ increases (from left to right), the relative distance between the observer and the photons decreases, causing the observed photon frequency to rise, and the blueshift effect begins to emerge. Consequently, the increase in photon energy makes the optical image brighter. Conversely, light moving away from the observer undergoes redshift, with a decrease in photon energy, resulting in a dimmer optical image. Thus, the observation inclination angle plays a dominant role in influencing the redshift and blueshift effects of a thin accretion disk.

\subsection{Influence of the Coupling Parameter $\Lambda$  on Optical Images}

\begin{figure}[H]
	\centering
	\includegraphics[scale=0.7]{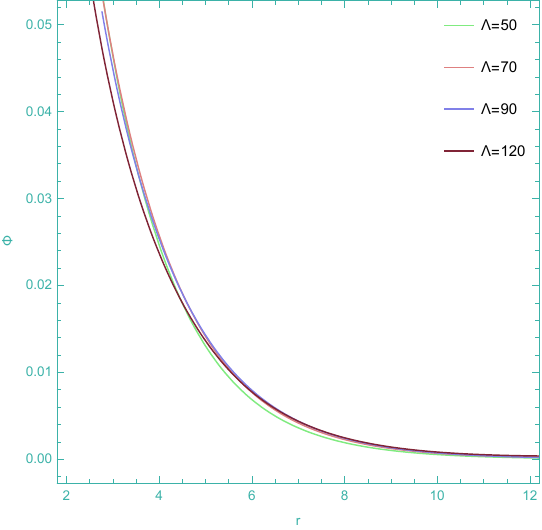}
	\caption{Variation of the scalar field $\Phi(r)$ with respect to the radial coordinate $r$ for different values of $\Lambda$. The values of $\Lambda$ considered are 50, 70, 90, and 120.}\label{fig8}
\end{figure}

\begin{figure}[H]
	\centering
	\includegraphics[scale=0.40]{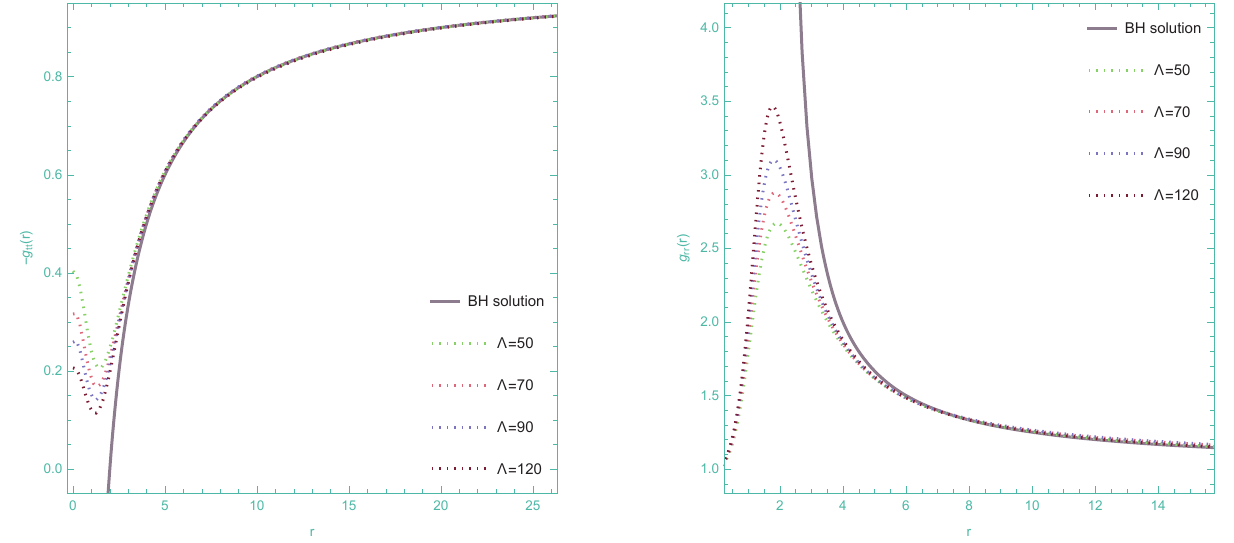}
	\caption{Comparison of the metric components $-g_{tt}$ (left panel) and $g_{rr}$ (right panel) between boson stars and the Schwarzschild black hole for different values of $\Lambda$. The dark solid lines indicate the Schwarzschild black hole, while the dashed lines represent boson stars. The values of $\Lambda$ considered are 50, 70, 90, and 120.}\label{fig9}
\end{figure}

In this section, the values of the magnetic charge $q$, free parameter $s$, and initial parameter $\phi_0$ are fixed at $q = 1$, $s = 1$, and $\phi_0 = 0.12$, respectively. For other values of these parameters, the results are similar. The numerical results of the scalar field $\Phi(r)$ for different values of the coupling constants $\Lambda$ are shown in Fig.~\ref{fig8}. The results indicate that the scalar field still exists only within a narrow range near $r=0$, which is consistent with all boson star models. The metric components $-g_{tt}$ and $g_{rr}$, for different values of $\Lambda$, are plotted in Fig.~\ref{fig9}, where the dark solid lines represent the metric components of the Schwarzschild black hole with the same mass parameter. The figure also reveals that the metric components of the Schwarzschild black hole diverge at the event horizon, whereas those of the boson stars remain finite. Their asymptotic behaviors converge as $r$ increases, with the metric components of both approaching 1.

\begin{figure}[H]
	\centering
	\includegraphics[scale=0.40]{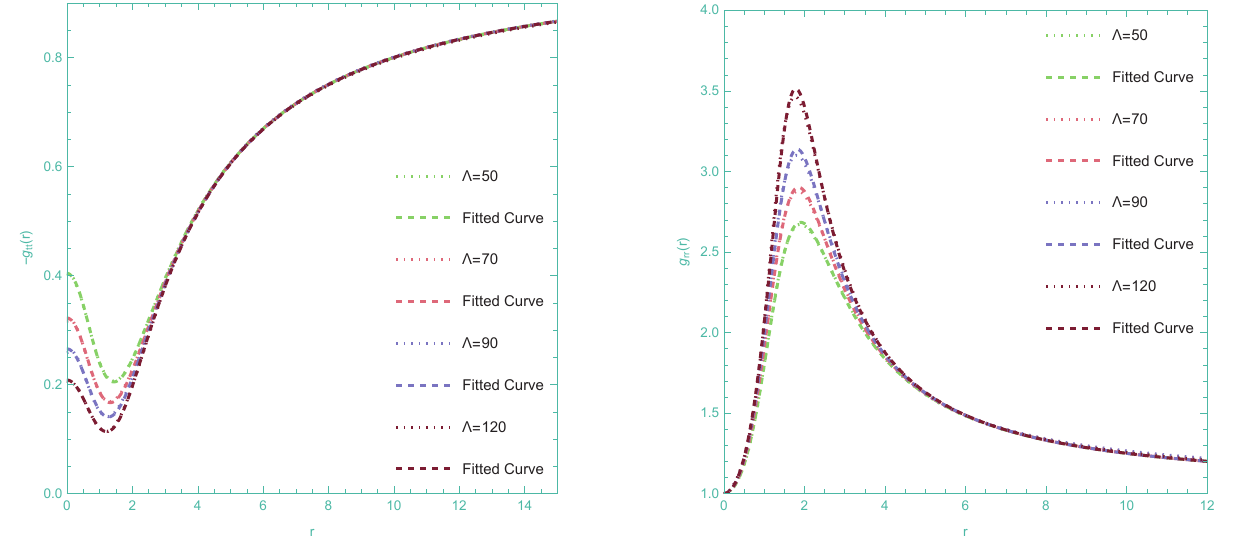}
	\caption{Comparison of the metric components $-g_{tt}$ (left panel) and $g_{rr}$ (right panel) between the numerical and fitted results for different values of $\Lambda$. The numerical results are depicted with the dotted lines, while the fitted results are depicted with the dashed lines. The values of $\Lambda$ considered are 50, 70, 90, 120.}\label{fig10}
\end{figure}

After obtaining the boson star solutions, we still use functions (\ref{function40}) and (\ref{function41}) to fit the metric components. Here, Fig.~\ref{fig10} presents the numerical and fitted results for the metric components. It can be observed that for the metric components $-g_{tt}$ and $g_{rr}$, the numerical results and the fitted curves are still identical. The parameters of the fitting functions, along with the boson star's ADM mass, are presented in Tables~\ref{tab4} and \ref{tab5}.

\begin{table}[H]
	\centering
	\caption{Values of $\alpha_i$ in the fitted metric component $-g_{tt}$ [see Eq.~(\ref{function40})] for different values of $\Lambda$ and $M$.}\label{tab4}
	\vspace{0.5mm}
	\begin{tabular}{ccccc}
		\hline
		Type&$M_{\Lambda}BS1$& $M_{\Lambda}BS2$ &$M_{\Lambda}BS3$  & $M_{\Lambda}BS4$  \\
		\hline
		$\Lambda$ &50  &70  &90 &120  \\
		$M$ & 0.783928  &0.823832 &0.843337&0.851462 \\
		$\alpha_1$&-0.506 &-0.081   &-0.105   &-0.671   \\
		$\alpha_2$&0.989  &1.074    &0.974    &0.673\\
		$\alpha_3$& -0.67 &-61764.8 &-99987.9 &-0.455 \\
		$\alpha_4$&  0.33 &-10640   &-19048   &0.293\\
		$\alpha_5$&-0.259 &38901.7  &67854    &-0.163 \\
		$\alpha_6$&-0.166 &-46344.3 &-78099.5 &-0.078 \\
		$\alpha_7$&-0.263 &-71113.1 &-135295  &-0.197 \\
		\hline
	\end{tabular}
\end{table}

\begin{table}[H]
	\centering
	\caption{Values of $\beta_i$ in the fitted metric component $g_{rr}$ [see Eq.~(\ref{function41})] for different values of $\Lambda$ and $M$.} \label{tab5}
	\vspace{0.5mm}
	\begin{tabular}{ccccc}
		\hline
		Type&$M_{\Lambda}BS1$& $M_{\Lambda}BS2$ &$M_{\Lambda}BS3$  & $M_{\Lambda}BS4$  \\
		\hline
		$\Lambda$ &50  &70  &90 &120  \\
		$M$ & 0.783928  &0.823832 &0.843337&0.851462 \\
		$\beta_1$&-20.613 &-18.373 &-16.335 &-14.34   \\
		$\beta_2$&-8.201  &-5.787  &-5.458 &-33.396 \\
		$\beta_3$&4.47    &4.836   &6.546  &499.953 \\
		$\beta_4$&0.511   &-0.238  &-2.074 &-307.66 \\
		$\beta_5$&4.12    &3.768   &4.527   &51.696 \\
		$\beta_6$&0.202   &0.145   &0.213  &122.84\\
		$\beta_7$&-0.042  &-0.059  &-0.102  &-6.791 \\
		\hline
	\end{tabular}
\end{table}

\begin{figure}[H]
	\centering
	\includegraphics[scale=0.55]{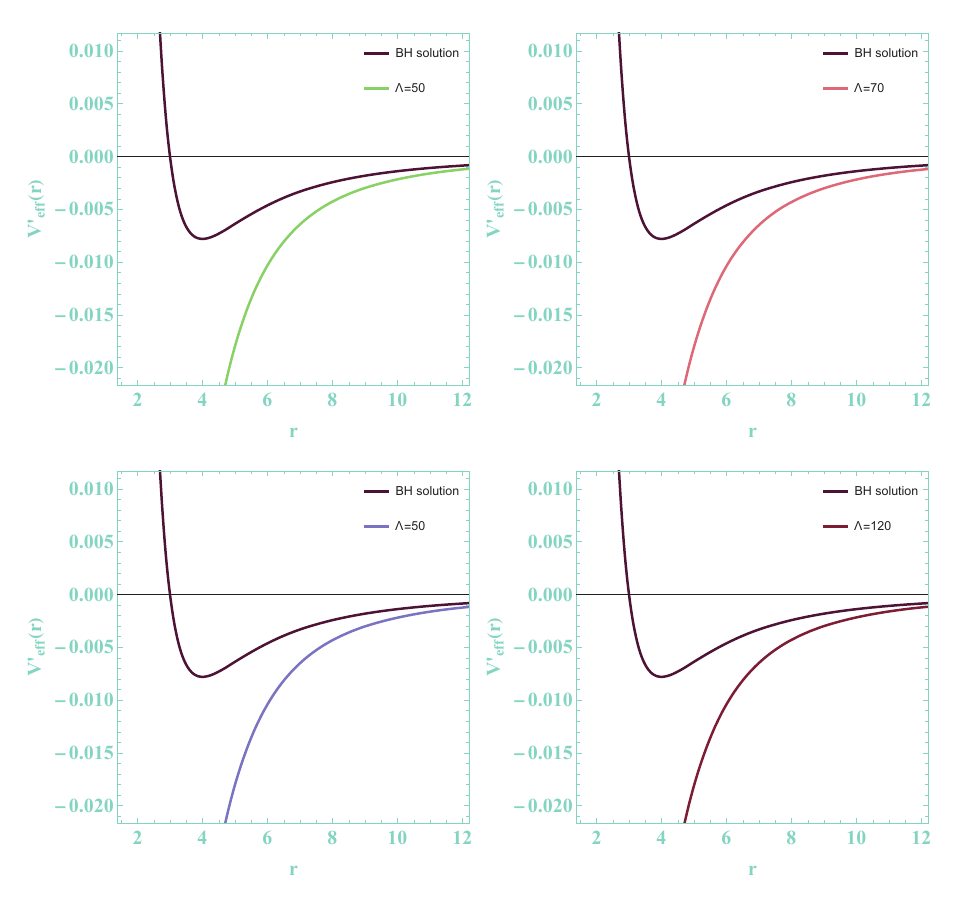}
	\caption{The first derivative of the effective potential for the boson stars ($M_{\phi_0}BS1-M_{\phi_0}BS4$) and the Schwarzschild black holes with the same mass parameters for different values of $\Lambda$. From left to right, in the first row $\Lambda=50$ and 70, and in the second row $\Lambda=90$ and 120.}\label{fig11}
\end{figure}

Similarly, with Eq.~(\ref{function27}), we plot the first derivative $V^{\prime}_{eff}(r)$ of the effective potential for the boson stars ($M_{\Lambda}BS1-M_{\Lambda}BS4$) in Fig.~\ref{fig11}. For comparison with black holes, each figure includes a dark solid line representing the first derivative of the effective potential for the Schwarzschild black hole with the same mass parameter. It is also found that  $V^{\prime}_{eff}(r)$ for the boson stars increases monotonically with the radial coordinate $r$. As $r\rightarrow \infty$, $V^{\prime}_{eff}(r)$ approaches zero without crossing the horizontal axis, indicating that the boson stars ($M_{\Lambda}BS1-M_{\Lambda}BS4$) lack photon rings. For all Schwarzschild black holes, a photon ring exists at $r=3M$.

\begin{figure}[H]
	\centering
	\includegraphics[scale=0.55]{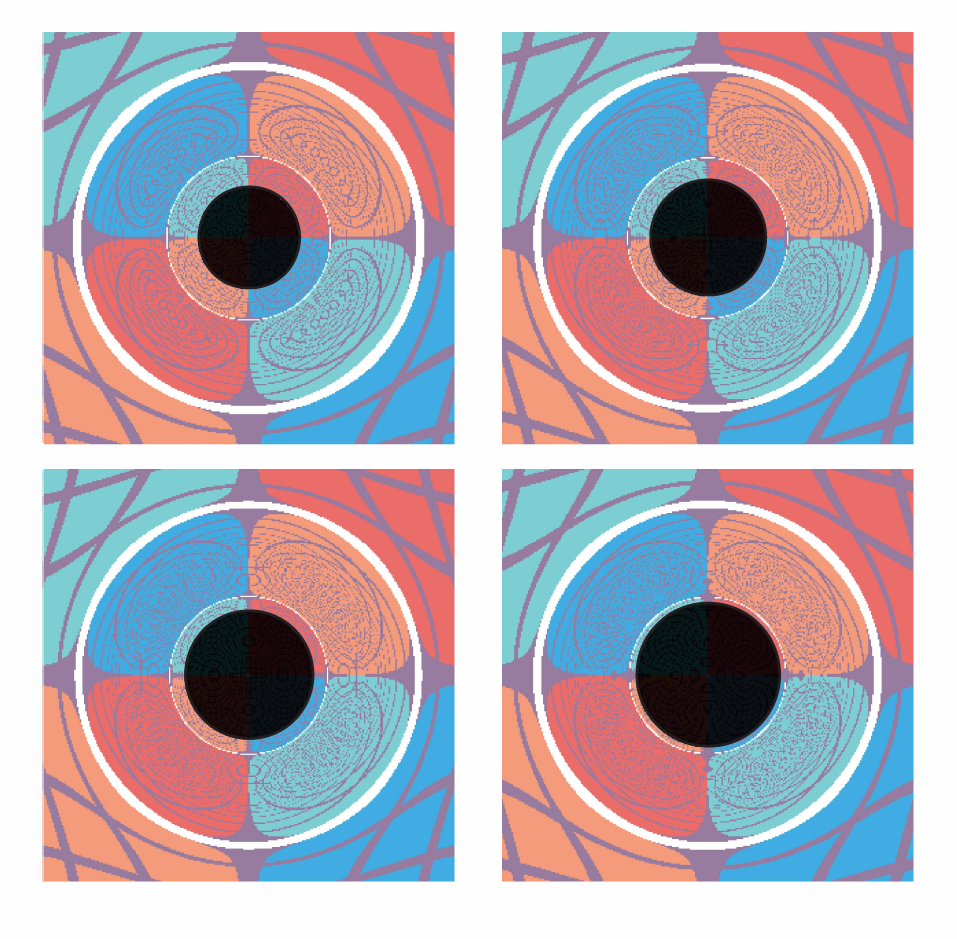}
	\caption{Optical images of boson stars and their Einstein rings observed at an observation inclination angle of $\theta=75^\circ$ with a field angle of $\gamma_{_{fov}}=18^\circ$ under celestial light source illumination. From left to right, in the first row $\Lambda=50$ and 70, and in the second row $\Lambda=90$ and 120. The magnetic charge $q$, free parameter $s$, and initial parameter $\phi_0$ are fixed at $q=1$, $s=1$, and $\phi_0=0.12$, respectively.}\label{fig12}
\end{figure}

In Fig.~\ref{fig12}, we plot the optical images of the boson stars ($M_{\Lambda}BS1-M_{\Lambda}BS4$) and their Einstein rings under celestial light source illumination. From these images, it can be observed that the primary features of the optical images and Einstein rings are consistent with those shown in Fig.~\ref{fig5}. Here, we focus on the influence of the coupling constant $\Lambda$ on the optical images of the boson stars. The results indicate that as $\Lambda$ increases, the radius of the boson star also increases, but the Einstein ring remains relatively unchanged, which is different from the case of the initial parameter $\phi_0$.

\begin{figure}[H]
	\centering
	\includegraphics[scale=0.4]{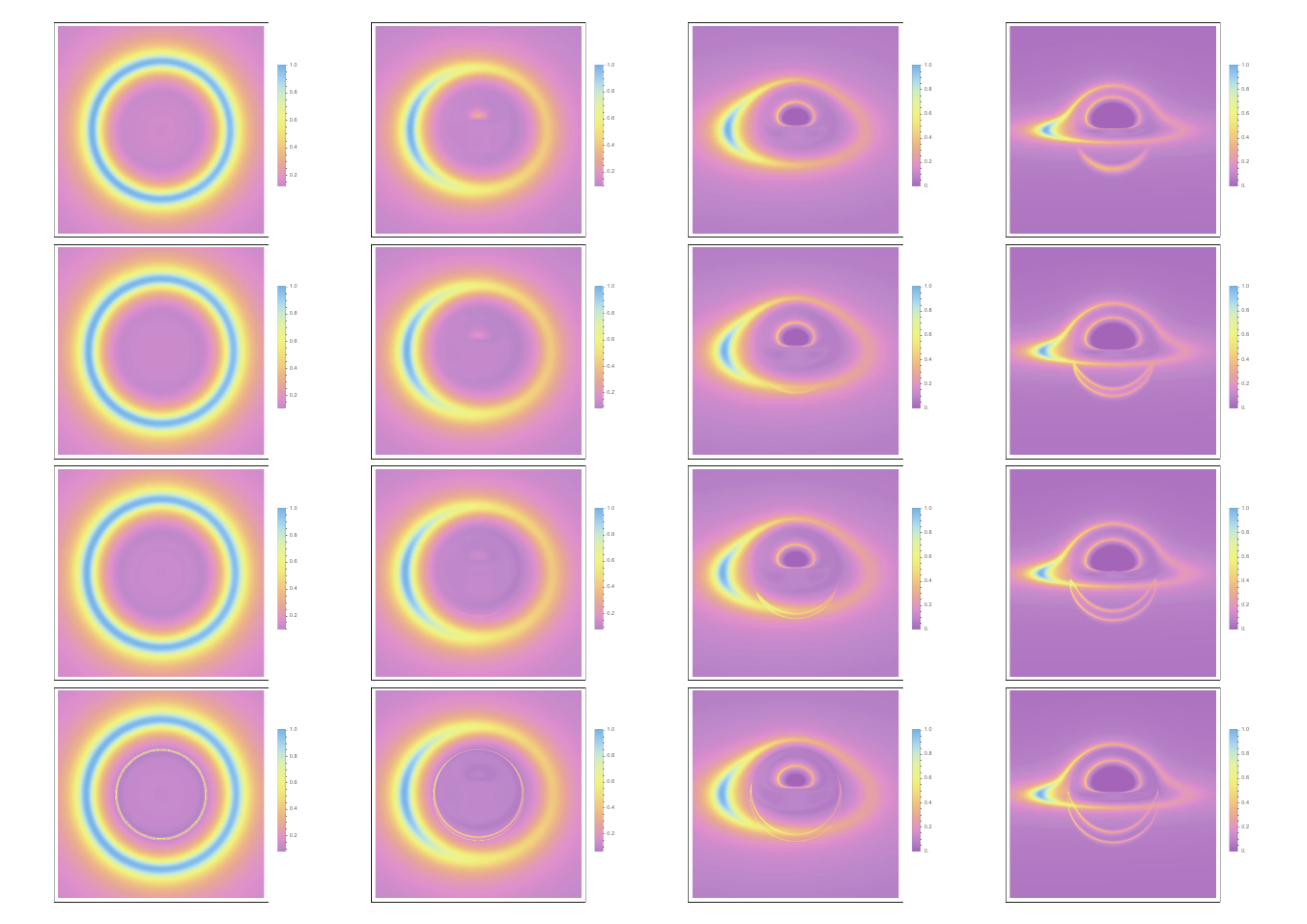}
	\caption{Optical images of boson stars displayed with a field angle $\gamma_{_{fov}}=5^\circ$ under thin accretion disk illumination. The vertical arrangement, from top to bottom, corresponds to $M_{\Lambda}BS1$ ($\Lambda=50$), $M_{\Lambda}BS2$ ($\Lambda=70$), $M_{\Lambda}BS3$ ($\Lambda=90$), and $M_{\Lambda}BS4$ ($\Lambda=120$), respectively. Horizontally, the values of the observation inclination angle $\theta$, from left to right, are $1^\circ$, $25^\circ$, $60^\circ$, and $80^\circ$, respectively. The magnetic charge $q$, free parameter $s$, and initial parameter $\phi_0$ are fixed at $q=1$, $s=1$, and  $\phi_0=0.12$, respectively.}\label{fig13}
\end{figure}

When considering a background light source modeled as a thin accretion disk, the images of the boson stars ($M_{\Lambda}BS1-M_{\Lambda}BS4$) are presented in Fig.~\ref{fig13}. The dark central region surrounded by the bright ring corresponds to the shape of the boson star. The images reveal that the coupling constant $\Lambda$ primarily affects the size of the direct image. As the observation inclination angle $\theta$ increases, the direct image transitions from a ``disk-like'' shape to a ``cap-like'' shape, with pronounced lensed images emerging. The size of these lensed images also grows with increasing $\theta$. In Fig.~\ref{fig11}, the first derivative of the effective potential for the boson stars ($M_{\Lambda}BS1-M_{\Lambda}BS4$) does not intersect the horizontal axis, indicating the absence of photon rings in these boson stars, which is also confirmed in Fig.~\ref{fig13}.

From the rows of Fig.~\ref{fig13}, it can be observed that the smaller value of $\Lambda$ results in minimal light deflection. As a result, the ``crescent-shaped'' lensed images are only visible at large observation inclination angles. In the fourth row of Fig.~\ref{fig13}, where $\Lambda$ is large enough, the light deflection is severe, leading to the following observations: (1) Lensed images are visible at any observation inclination angle. (2) The size of the lensed images increases with the observation inclination angle, while their shape transitions from ``ring-like'' to ``crescent-like''. (3) Two distinct ``ring-like'' structures can be seen, corresponding to two components of the lensed image. The first component arises from photons orbiting the boson star beyond half of the equatorial plane, a feature also observed in black hole shadow images. The second component results from photons plunging directly through the interior of the boson star (referred to as plunge-through orbit). In black hole spacetimes, this component is absent, as photons cannot escape from within the event horizon. These phenomena demonstrate that the coupling constant $\Lambda$ primarily influences the size of the direct image, whereas the observation inclination angle $\theta$ primarily influences the shape of the direct image and the size of the lensed image. As $\theta$ increases, the peak light intensity shifts toward the left side of the image, while the intensity on the right side gradually diminishes.

\begin{figure}[H]
	\centering
	\includegraphics[scale=0.4]{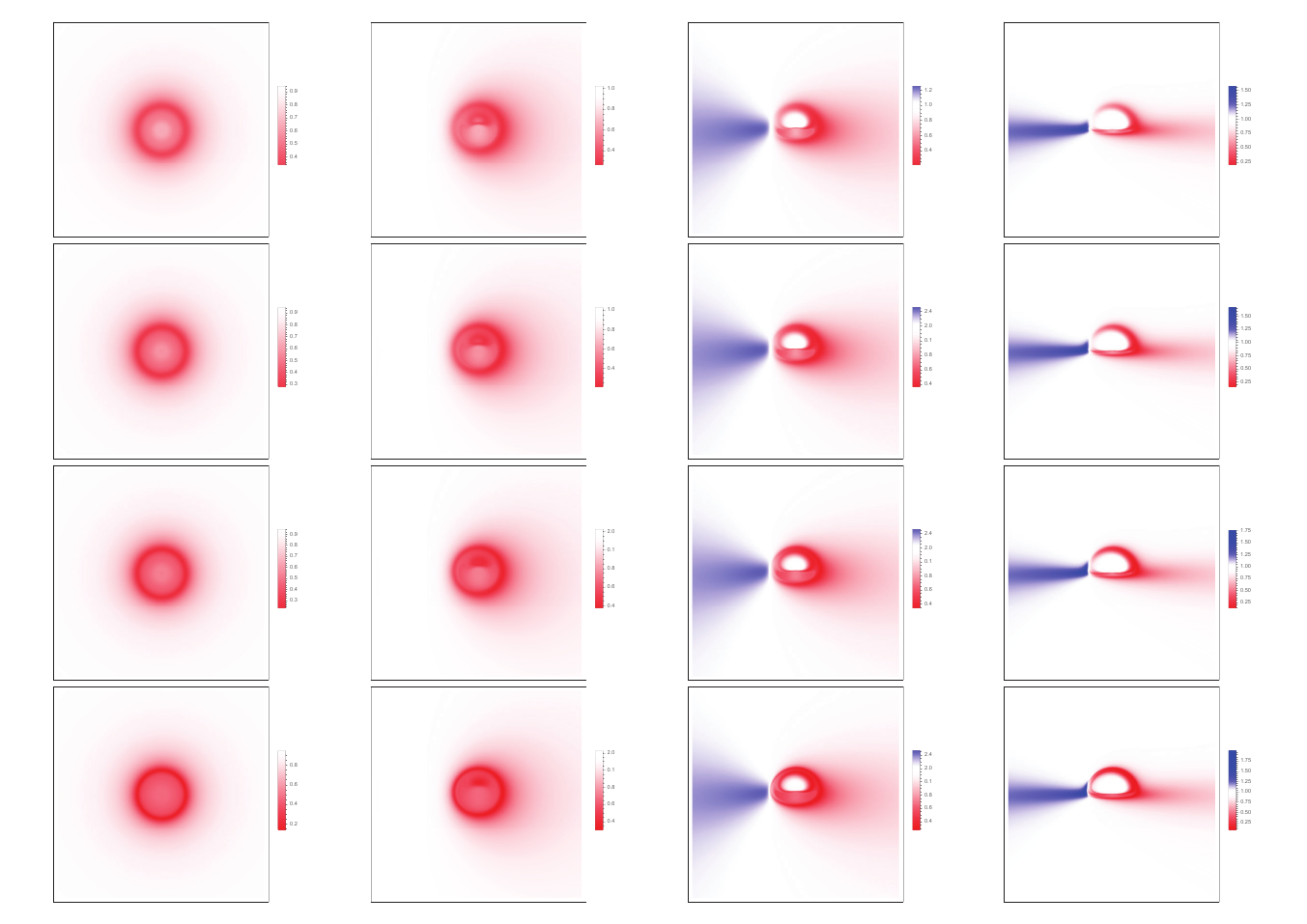}
	\caption{Optical images of boson stars displayed with a field angle $\gamma_{_{fov}}=10^\circ$ under thin accretion disk illumination. The vertical arrangement, from top to bottom, corresponds to $M_{\Lambda}BS1$ ($\Lambda=50$), $M_{\Lambda}BS2$ ($\Lambda=70$), $M_{\Lambda}BS3$ ($\Lambda=90$), and $M_{\Lambda}BS4$ ($\Lambda=120$), respectively. Horizontally, the values of the observation inclination angle $\theta$, from left to right, are $1^\circ$, $25^\circ$, $60^\circ$, and $80^\circ$, respectively. The magnetic charge $q$, free parameter $s$, and initial parameter $\phi_0$ are fixed at $q=1$, $s=1$, and $\phi_0=0.12$, respectively.}\label{fig14}
\end{figure}

The images illustrating the redshift and blueshift effects of a thin accretion disk are shown in Fig.~\ref{fig14}. Similar to the previous case, as $\theta$ increases, a blueshift emerges. According to Eq.~(\ref{function35}), when $\theta=1^\circ$, the direct image exhibits redshift, primarily due to the gravitational redshift. As $\theta$ increases, the blueshift effect begins to appear and gradually intensifies. For large observation inclination angles, the observed redshift effect is mainly attributed to Doppler redshift.

\section{Conclusion}

This study has numerically derived static, spherically symmetric solutions for massive boson stars within a theoretical framework that integrates Einstein's nonlinear electrodynamics and a complex scalar field minimally coupled to gravity. Using ray-tracing simulations, we systematically investigated the optical images of the boson stars under both celestial light source illumination and thin accretion disk illumination.

The boson star solutions yield asymptotically flat spacetimes without the event horizon, as demonstrated by the finite behavior of the metric components $-g_{tt}$ and $g_{rr}$ (see Figs.~\ref{fig2} and \ref{fig9}). Although they share the asymptotic behavior of Schwarzschild black holes, the lack of the event horizon fundamentally alters photon dynamics and observational signatures. The first derivative of the photon's effective potential for the boson stars displays a monotonically increasing behavior with respect to the radial coordinate, and does not intersect the horizontal axis (see Figs.~\ref{fig4} and \ref{fig11}). This phenomenon indicates that photon rings are absent for the boson stars. Notably, the absence of the event horizon allows photons from the boson star's interior to escape, resulting in a distinctive central bright region (see Figs.~\ref{fig5} and \ref{fig12}). For thin accretion disk illumination, the direct emission component dominates at small observation inclination angles, while lensed emission components become prominent for large observation inclination angles (see Figs.~\ref{fig6} and \ref{fig13}). Additionally, the unique ``plunge-through orbit'' photons traversing the stellar interior become observable (see Figs.~\ref{fig6} and~\ref{fig13}), a phenomenon not possible for black holes. For both illuminations, the redshift/blueshift distributions exhibit strong $\theta$ dependence: the redshift effect prevails at small $\theta$, while the blueshift effect appears at large $\theta$ (see Figs.~\ref{fig7} and \ref{fig14}).

Regarding the initial parameter $\phi_0$ and coupling parameter $\Lambda$, we have the following conclusions. Increasing $\phi_0$ enlarges the stellar radius, resulting in a larger Einstein ring under celestial light source illumination (see Fig.~\ref{fig5}) and a diminished bright ring structure under thin accretion disk illumination (see Fig.~\ref{fig6}). Increasing $\Lambda$ dose not significantly affect the Einstein ring under celestial light source illumination (see Fig.~\ref{fig6}), and primarily reduces the size of the direct image under thin accretion disk illumination (see Fig.~\ref{fig13}).

These findings provide robust theoretical predictions for the optical signatures of massive boson stars with nonlinear electrodynamics. Therefore, this study may lay a theoretical foundation for interpreting high-resolution imaging data from next-generation instruments, such as the ngEHT, facilitating the observational differentiation between black holes and massive boson stars with nonlinear electrodynamics.

\vspace{10pt}

\noindent {\bf Acknowledgments}\\

\noindent {This work is supported by the National Natural Science Foundation of China (Grant No. 12375043), the Natural Science Foundation of Chongqing (CSTB2023NSCQ-MSX0594), and the China Postdoctoral Science Foundation (Grant No. 2024M753825).}

\end{document}